\documentclass[reprint,prb,amsmath,amssymb,aps]{revtex4-2}
\usepackage{silence}
\usepackage[utf8]{inputenc}
\usepackage{graphicx}
\usepackage{amsmath,amssymb}
\usepackage{dsfont}
\usepackage{xcolor}
\usepackage{amsthm}
\usepackage{float}
\usepackage{appendix}
\usepackage{dcolumn}
\usepackage{bm}
\usepackage{hyperref}
\usepackage{braket}

\begin{document}

\preprint{APS/123-QED}

\title{
Fractalizing spacetime: Floquet codes with fractonic excitations that are immobile in space and time}

\author{Juliette Soule}
\author{Dominic J.~Williamson}
\affiliation{School of Physics, The University of Sydney, Sydney, NSW 2006, Australia}

\date{Spetmber 2026}

\begin{abstract}
\noindent
We generalize fractalization, a procedure for the construction of fracton models, from space to spacetime. 
We apply spacetime fractalization to construct fracton floquet codes with syndrome excitations that have limited mobility in space and time. 
This extends the notion of fracton order to intrinsically dynamical quantum phases of matter that are inequivalent to static fracton phases. 
We find spacetime type-II fracton floquet codes which have no topological excitations that are mobile in space or time. These codes exhibit an extreme form of quantum discrete time crystal order with response periods that scale exponentially in their linear system sizes. 
In this context, the no-strings rule that characterizes type-II fractons leads to a superlinear scaling of the floquet code fault-distance with time, potentially lowering the time overhead required for quantum error~correction. 
\end{abstract}
\maketitle

\section{Introduction}\label{intro}
Quantum spin models with fractal symmetries have been a topic of significant interest to research communities in both condensed matter~\cite{Yoshida_2013,fractongauge,fracton7,Devakul_2019} and quantum information theory~\cite{Haah_2011,Brown_2016,quantum1,quantum2,Devakul_2018,Miguel_2023}. These models exhibit topological order beyond the description of conventional topological quantum field theory~\cite{Aasen2020,Song2023}. 
The canonical example of such a model is Haah's Cubic Code \cite{Haah_2011}. The Cubic Code possesses topological order with stability against local perturbations, however it is distinct from models with conventional topological order. In particular, the model satisfies a no-strings rule and consequently has immobile `fracton' excitations which are created at the corners of logical operators with fractal support. The cubic code is characterized as having Type-II Fracton topological order~\cite{fracton2,fracton3,fracton4,fracton5,fracton6,fracton8,Williamson_2021_TypeII}.

Type-II fracton topological orders are inherently interesting due to their unique physical characteristics amongst quantum phase of matter. These physical characteristics imbue Type-II fracton models with great potential as quantum memories. Type-II fracton models are defined by their lack of stringlike logical operators, which is intimately related to their immobile quasiparticle excitations. The Cubic Code, in particular, has been shown to exhibit partial self-correction at finite temperature~\cite{quantum2}. 

Fractalization provides a lens through which quantum spin models with fractal symmetry can be viewed~\cite{Yoshida_2013,dom_main}. The fractalization procedure maps operators on a $D$ dimensional lattice to a $D+m$ dimensional lattice, via a linear cellular automaton rule of dimension $m$. Fractalized codes are characterised by logical operators with fractal support, and in turn immobile fracton excitations created by truncations of these logical operators. Many quantum spin models with fractal symmetries, including fracton topological orders, can be understood as fractalized versions of simpler, lower dimensional codes. For instance, the Cubic Code is equivalent to a fractalized 2D Toric Code~\cite{Kitaev_2003}. 

In this work, we extend the framework of fractalization to spacetime, thus generating new families of Floquet codes that possess fractal symmetries in both space and time.  In this way, we find codes that can serve as fault-tolerant quantum memories with a logical evolution period that is exponentially long in the system size. This is achieved via fractal structures that are naturally generated by linear cellular automata~\cite{lca1,lca2,lca3,lca4}, which themselves have exponentially long periods~\cite{Martin1984}. By fractalizing spacetime we construct floquet codes with fracton excitations that are immobile in both space and time. This generalizes the notion of a Type-II fracton, satisfying the no-strings rule, from space to spacetime. 

\begin{figure}[t]
    \centering
    \includegraphics[scale=0.66,page=1]{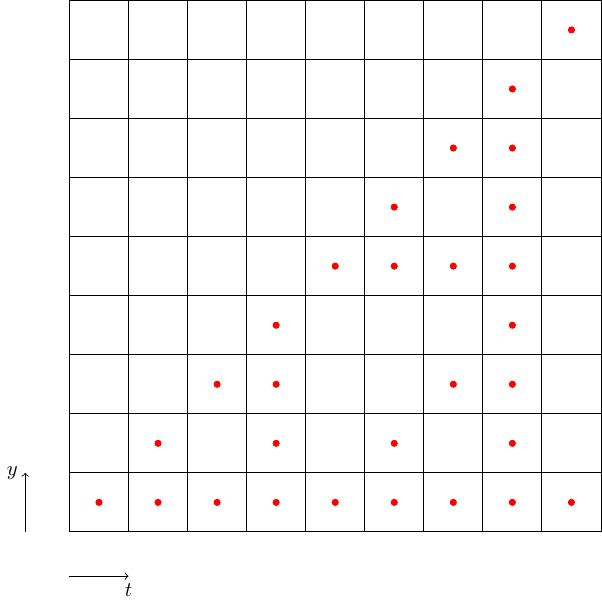}
    \caption{Discrete fractal pattern generated over time by the Sierpinski Linear Cellular Automaton rule, $f(y)=1+y$. }
    \label{fig:sierpinski}
\end{figure}

There has been a large volume of literature on Floquet codes in (2+1)D and (3+1)D~\cite{Hastings2021,Paetznick2022,Vuillot2021,Gidney2021,Haah2022,Aasen2022,zhang2022xcubefloquetcode,Ellison2023,Aasen2023,Davydova2023,Kesselring2024,GransSamuelsson2024,Davydova2024QAC,Claes2025,Setiawan2025,williamson2025dynamicalquantumcodeslogic}. Despite the discovery of many new floquet codes, including some with period doubling~\cite{Hastings2021,aasen2023measurementquantumcellularautomata}, the topological phasew realized by these floquet codes in spacetime are all equivalent to those realized by conventional static codes (when time translation symmetry is not enforced)~\cite{williamson2023spacetime}. 
Our work applies fractalization in time to go beyond existing examples and construct floquet codes that realize new topological phases in spacetime, beyond any static counterpart. This extends the application of fractalization in space to generate topological orders that go beyond a conventional TQFT description~\cite{Aasen2020,Song2023}. 

To achieve our results, we use the framework of measurement based quantum computing (MBQC) to construct fractalized floquet codes. MBQC is an established framework in which resource states for quantum computing via measurements along furnish a duality between space and time. A wide variety of states have been shown to be useful as resources for MBQC~\cite{mbqc1,mbqc2,mbqc3,mbqc4,mbqc5,mbqc6,mbqc7,mbqc8,mbqc9,mbqc10,mbqc11}. MBQC is a natural bridge to generalize the fractalization of space~\cite{dom_main} to spacetime.

\subsection*{Summary of Results}

\begin{figure}[t]
    \centering
   \includegraphics[page=3,scale=0.85]{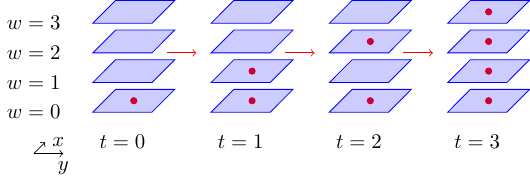}
    \caption{Evolution of an excitation in the time-fractalized toric code (layers labelled by $w$). With each timestep $t$ (left to right) the excitation (red circle) propagates under the application of the linear cellular automaton update rule $h(w)=1+w$, which is indicated via a red arrow.}
    \label{fig:toric2d}
\end{figure}

\begin{figure}[t]
    \includegraphics[page=26,scale=0.85]{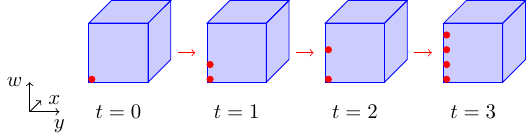}
    \caption{Evolution of an excitation in a spacetime type-II floquet code. With each timestep $t$ (left to right) the fracton (red circle) propagates under the application of the linear cellular automaton update rule $h(w)$, which is represented by the red arrow.}
    \label{fig:spacetimeTII}
\end{figure}

Our main result is the construction of an intrinsically \textit{spacetime type-II} fracton floquet code. 
There are several steps to achieve this result. 
First, we derive constant-depth local adaptive quantum circuit implementations of linear cellular automata evolutions, such as the Sierpinski rule in Figure~\ref{fig:sierpinski}, see Section~\ref{circuitimplementation}. 
Next, we construct a fault-tolerant implementation of these linear cellular automata evolutions using transversal constant-depth local adaptive quantum circuits across a stack of quantum error-correcting code blocks, such as the Toric Code, see Section~\ref{concatenation}. 
This procedure implements fractalization in the time direction of the quantum memory based on a single code block from the stack, see Figure~\ref{fig:toric2d}. 
Finally, we apply fractalization in space and time to the (2+1)D Toric Code quantum memory to produce a (3+1)D spactime type-II floquet code in which a type-II Fractal Spin Liquid model~\cite{Yoshida_2013} evolves under an independent local cellular automaton rule, see Figure~\ref{fig:spacetimeTII}, resulting in quasiparticle excitations that have no string operators in spacetime, see Section~\ref{logicals}. 

\subsection*{Outline}

The manuscript is laid out as follows. 
In Section~\ref{sec:Pre} we provide preliminary material about fractals generated by linear cellular automata. 
In Section~\ref{sec:Sierpinski} we introduce a constant-depth local adaptive circuit realization of linear cellular automata. 
In Section~\ref{concatenation} we apply this circuit transversally to a stack of quantum error-correcting codes to define the \textit{fractalization} of time for the quantum memory based on a single code. 
In Section~\ref{sec:Excitations} we show how fractalization leads to immobile excitations in the space and time directions. 
In Section~\ref{sec:disc} we discuss our results.

\section{Background: Linear Cellular Automata and Fractals}
\label{sec:Pre}

In this section we review linear cellular automata (LCA) and their connection to discrete fractals. Throughout this work our description of dynamical fracton codes generated by fractalization in space and time is based on classical LCA and the discrete fractal patterns that they generate.
We take advantage of the fact that logical operators of a fractalized code inherit the fractal history of an underlying LCA. Similarly, the immobile fracton excitations in these codes are created by truncating the logical operators.

We consider spins at sites along an infinite 1D chain.  Each site $i$ is associated to the variable $a_{i}\in\{0,1\}.$ The state of an LCA at time $t$ is specified by the binary string $\{a_{i}^{t}\}.$ The dynamics of an LCA is defined by linear update rules, which determine the state at the next time step $\{a_{i}^{t+1}\}$ from the state at the present time step $\{a_{i}^{t}\}.$ We focus on the Sierpinski LCA to illustrate the general formalism. For the Sierpinski LCA, the update rule is
\begin{align}
    a_{i}^{t+1} = a_{i}^{t}+a_{i-1}^{t}.
\end{align}
It is convenient to describe LCA in the polynomial representation, where the state $\{a_{i}^{t}\}$ is encoded in the Laurent polynomial
\begin{align}
    s_{t}(x) &=\sum_{i=-\infty}^{\infty}a_{i}^{(t)}x^{i}.
\end{align}
Update rules now take the form $s_{t+1}(x) = f(x)s_{t}(x)$ for some polynomial $f(x).$ For the  Sierpinski LCA, $f(x) = 1+ x.$ 
Powers of the generating polynomial
\begin{align}
    f^{0} &= 1\\ \nonumber
    f^{1} &= 1+x\\\nonumber
    f^{2} &= 1\quad\;\;\;+x^{2}\\\nonumber
    f^{3} &= 1+x+x^{2}+x^{3}\\\nonumber
    f^{4} &= 1\quad\quad\quad\quad\quad\quad+x^{4}\nonumber
\end{align}
give rise to a discrete Sierpinski triangle fractal. The entire Sierpinski triangle can be represented by the bivariate polynomial
\begin{align}
    g(x,y)&=1+f(x)y+f^{2}(x)y^{2}+f^{3}(x)y^{3}+...
\end{align}
The Sierpinski LCA generates a discrete fractal structure seeded by $s_{0}(x) = 1.$ 
This follows from the fact that the polynomials in $\mathbb{F}_{2}$ satisfy $f(x)^{2^{n}}=f(x^{2^{n}})$. 
Consider an LCA acting on the initial state $s_{0}(x) = 1,$ ie $a_{0} = 1, a_{i} = 0$, for all $ i\neq 1.$ At a later time $t = 2^{n}$, we have $s_{t}(x) = f(x)^{2^n} = x^{2^{n}}+1,$ ie $a_{0} = 1, a_{2^{n}} = 1.$ Therefore, the state at $t = 2^{n}$ is the initial state, plus a copy of the initial state shifted by $2^{n}.$ At subsequent times, this process will repeat itself at a larger scale. 
More generally, the fractal structures generated by LCA on finite chains with periodic boundary conditions display periodic behaviour: After a possible `transient' phase, such an LCA must enter a cycle in which it revisits the same configurations periodically~\cite{Devakul_2018}. 
If we have $N$ sites and periodic boundary conditions, the cycle length as a function of $N$ is within an exponentially growing envelope which saturates periodically~\cite{LCA}. 

The discussion above focused on LCA generated dynamics on classical spins. The extension of such LCA dynamics to a system of quantum spins is straightforward as we now describe. 
This extension is based on the fractalization procedure for classical and quantum codes~\cite{dom_main}. 
In order to fractalize stabilizer code Hamiltonians, it is convenient to represent the Hamiltonian interaction terms, which are $X$ or $Z$-type Pauli operators, by polynomials~\cite{Yoshida_2013,Haah2013}. We follow the notation used in Ref.~\cite{dom_main} and consider the sites $\textbf{r}$ of a lattice and the Pauli operators $X_{\textbf{r}},Z_{\textbf{r}}$ acting on those sites. We define a map $\sigma_{X},\sigma_{Z}$ from the polynomials $a(x) = \sum_{\textbf{r}}a_{\textbf{r}}x^{\textbf{r}}, b(x) = \sum_{\textbf{r}}b_{\textbf{r}}x^{\textbf{r}}$ to a tensor product of Paulis as follows
\begin{align}
    \sigma_{X}[a(x)]&=\prod_{\textbf{r}}X_{\textbf{r}}a^{\textbf{r}}\nonumber\\
    \sigma_{Z}[b(x)]&=\prod_{\textbf{r}}Z_{\textbf{r}}b^{\textbf{r}}.
\end{align}
It is straightforward to express the action of translations in space in this notation: translation of the operator $\sigma_{X}[a(x)]$ by $\textbf{r}$ is implemented by multiplying the input polynomial by $x^{\textbf{r}}$ to obtain $\sigma_{X}[x^{\textbf{r}}a(x)].$
In order to describe the commutation relations between arbitrary translations of operators $\sigma_{X}[a(x)], \sigma_{Z}[b(x)]$ we introduce the spatial inversion operation $\bar{x} = x^{-1}.$ Then 
\begin{align}
    [\sigma_{X}[x^{\textbf{r}}a(x)],\sigma_{Z}[x^{\textbf{r}'}b(x)]]=(-1)^{c_{\textbf{r}-\textbf{r}'}}
\end{align}
where the polynomial that encodes the commutation relations is
\begin{align}
    c(x) = \sum_{\textbf{r}}x^{\textbf{r}}c_{\textbf{r}}=a(x)b(\bar{x}). 
\end{align}
The special case where $c(x) = 0$ implies that all translations of the operators considered commute. 

The models considered in this work may have more than one physical qubit per lattice site. In order to describe this, the above notation is adapted to include a column of $N$ polynomials and output a column of $N$ tensor products of Pauli operators, for a lattice site with $N$ physical qubits~\cite{Haah2013}.

\section{Sierpinski Cluster State}
\label{sec:Sierpinski}

In this section we introduce a procedure for fractalization of the time dimension for a quantum memory based on a general CSS stabilizer code~\cite{Calderbank1996,Steane1996}. 
We illustrate this procedure via the simple example of a fractalized CSS 1D cluster state~\cite{dom_main} which is equivalent to the Sierpinski cluster state. By analyzing this model we establish a framework for fractalizing the time dimension of a quantum memory which extends to higher dimensional examples. We describe an adaptive circuit to implement the time step of the  time-fractalized code. In the following section this is generalised to a fault tolerant setting.

\subsection{Fractalizing the 1D CSS Cluster State}\label{Vformalism}
We begin by considering the 1D CSS cluster state, which is a spin model with two qubits $a,b$ per site, and Hamiltonian 
\begin{align}
    \mathcal{H}&=-\sum_{\textbf{r}}\sigma_{X}\left[x^{\textbf{r}}\binom{1+x}{1}\right]+\sigma_{Z}\left[x^{\textbf{r}}\binom{1}{1+\bar{x}}\right]\nonumber\\
    &=-\sum_{\textbf{r}}X^{a}_{\textbf{r}}X^{b}_{\textbf{r}}X^{a}_{\textbf{r}+1}+Z^{b}_{\textbf{r}-1}Z^{a}_{\textbf{r}}Z^{b}_{\textbf{r}}.
\end{align}
This Hamiltonian energetically enforces a history groundstate corresponding to the trivial evolution of a qubit along a wire on the $a$ sublattice, up to sign choices given by spins on the $b$ sublattice~\cite{Raussendorf2001}.
Following the protocol in Ref.~\cite{dom_main}, we obtain the fractalized Hamiltonian as follows
\begin{align}\label{fracH}
    \mathcal{H} &=-\sum_{\textbf{r}}\sigma_{X} \left[x^{\textbf{r}}y^{s}\binom{1+f(y)x}{1} \right] +\sigma_{Z} \left[x^{\textbf{r}}y^{s}\binom{1}{1+\bar{f}(\bar{y})\bar{x}}\right]\nonumber\\
     &=-\sum_{\textbf{r}}\sigma_{X} \left[x^{\textbf{r}}y^{s}\textbf{A}(f(y) x) \right] +\sigma_{Z} \left[x^{\textbf{r}}y^{s}\textbf{B}(f(\bar{y}) \bar{x}) \right]
\end{align}
where
\begin{align}
    \textbf{A}(x)&=
    \begin{bmatrix}
        1 + x\\
        1
    \end{bmatrix},\quad
    \textbf{B}(x)=
    \begin{bmatrix}
    1\\
    1+x
    \end{bmatrix}.
\end{align}
The Hamiltonian in Eq.~\ref{fracH} enforces that the history ground state obeys the LCA evolution rule $f(y)$ in the $Z$ basis on the $a$ sublattice. 
Here, the $b$ sublattice spins can be thought of as providing choices of sign for the local constraints of the LCA. 

\begin{figure}[t]
    \centering
    \includegraphics[page=28,scale=1.5]{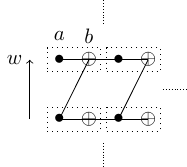}
    \caption{A segment of the Sierpinski cluster state. Layers extend vertically in the $w$ dimension (dotted lines). Qubits at lattice sites $a$ are measured in the $X$ basis and at sites $b$ are measured in the $Z$ basis.}
    \label{fig:sierpinski1d}
\end{figure}

Taking $f(y) = 1+y,$ we arrive at the Hamiltonian for the Sierpinski cluster state~\cite{kubica2018ungauging,Devakul_2019}
\begin{align}
    \mathcal{H} = -\sum_{i,j}Z^{b}_{i,j}Z^{b}_{i,j-1}Z^{b}_{i-1,j-1}Z^{a}_{ij} + X_{ij}^{a}X^{a}_{i,j+1}X_{i+1,j+1}^{a}X_{ij}^{b}.
\end{align} 
Fig.~\ref{fig:sierpinski1d} depicts four unit cells of the Sierpinski cluster state, where a unit cell is a site $ij$ with subsites $a,b$. The site $i(j+1)$ is shifted along the positive horizontal ($x$) direction, and the site $(i+1)j$ is shifted along the positive vertical  ($w$) direction. 

\subsubsection{1D Cluster State}
We now review the effect of measurement on the standard 1D cluster state, before moving on to describe the effect of measurement on the Sierpinski cluster state. 
We consider the initial product state $$|\psi\rangle_{0}\otimes|+\rangle_{1}\otimes|0\rangle_{2}\otimes|+\rangle_{3} \otimes|0\rangle_{4}\cdots\otimes|+\rangle_{N-1}\otimes|0\rangle_{N}.$$ 
The 1D cluster state is formed by applying $CX$ gates $CX_{i,i-1}CX_{i,i+1}$ for $i=1,3,..,N-1$. 
The 1D cluster state allows the logical state $|\psi\rangle$ to be teleported from qubit $i=0$ to qubit $i=N.$ This is accomplished by measuring each qubit of the cluster state except qubit $N.$ The measurement basis alternates, qubits initialised in $|+\rangle$ are measured in the X basis, while qubits initialised in $|0\rangle$ are measured in the Z basis. When the logical state is stored on qubit $i-1$, measuring qubit $i-1$ in the correct basis teleports the state to qubit $i,$ up to a Pauli operator that is determined by the measurement outcome. Keeping track of the Pauli frame via the measurement outcomes enables an appropriate Pauli byproduct operator to be applied at to qubit $N$, thus recovering the initial logical state $|\psi\rangle$~\cite{Briegel2009}.

Consider a 1D lattice consisting of $N$ sites, each containing two sublattice sites, $a$ and $b$ where a qubit is located. The qubit at $a$ is initialised in the $|+\rangle$ state and the qubit at $b$ is initialised in the $|0\rangle$ state. The qubits at site $i$ are entangled by a unitary $U_{ab}^{(i)}=CX_{a_{i}b_{i}}$. Suppose that the qubit at subsite $a$ is in the state $|\psi\rangle,$ 
then we have
\begin{align}
    CX_{ab}(|\psi\rangle_{a}|0\rangle_{b})&=\frac{1}{\sqrt{2}}(|+\rangle_{a}|\psi\rangle_{b}+|-\rangle_{a}Z|\psi\rangle_{b}).
\end{align}
We see that entangling $a$ with $b$ then measuring $a$ in the $X$ basis teleports the state $|\psi\rangle$ from $a$ to $b$ up to a $Z$ byproduct operator determined by the outcome of the $X$ measurement. For a full cluster with $N$ sites this procedure of entanglement and measurement is described by the operators
\begin{align}
    U&=\prod_{i=1}^{N}CX_{a_{i}b_{i}}CX_{a_{i+1}b_{i}}\\
    \mathcal{M}&=\prod_{i=1}^{N}M_{X_{a_i}}M_{Z_{b_i}}.
\end{align}

\subsubsection{Fractalized 1D Cluster State}
We now analyse the effect of measurements on the fractalized cluster state following a similar strategy to the previous subsection. 
We consider a fractalized cluster state with layers along the $w$ direction, and sites $i$ along the $x$ dimension which represents time. 
There are 2 qubits $a$ and $b$ located at each site. 
For the Sierpisnki example $f(w)=1+w$ the effect of measuring the cluster state to propagate the quantum state along the $x$ direction can be derived from the measurement of sites $a,b$ respectively. 
For the 1D cluster state, the effect of measuring successive qubits is to teleport a logical state along the chain. In the Sierpinski example, measuring the $a$ sublattice qubits on columns of sites along the $w$ direction with fixed $x$ coordinate teleports the logical information to the next site and while implementing the fractalization of time between layers, corresponding to the LCA rule. Measuring the $b$ qubits on sites in a column along the $w$ direction, in contrast, simply teleports the logical information along the chain. Specifically for sites $ij$ and sublattices $a,b,$  we can express the entangler circuit that creates the Sierpinski cluster as 
\begin{align}\label{eq13}
    \mathcal{U}=\prod_{ij}CX_{a_{ij} b_{ij}} CX_{a_{ij}b_{i(j+1)}} CX_{a_{(i+1)j} b_{ij}}
\end{align} 
which is subject to the following single qubit measurements 
\begin{align}\label{eq14}
    \mathcal{M} = \prod_{ij}\mathcal{M}{X_{a_{ij}}}\mathcal{M}{Z_{b_{ij}}} .
\end{align}
More generally, for an arbitrary LCA $f(w)$, let the number of terms in $f(w)$ be $n_{f}+1,$ i.e.~${n_{(1+w+w^3)}=2}$. We abuse notation and use $f^{(i)}(w)$ to describe the location of those terms, i.e.~${(1+w+w^3)^{(0)}=0},$ ${(1+w+w^3)^{(1)}=1},$ $(1+w+w^3)^{(2)}=3$. 
We can express the general fractalized cluster state entangler as 
\begin{align}
    &\mathcal{U}=\prod_{ij} \big( CX_{a_{(i+1)j}b_{ij}} \prod_{z=0}^{n_{f}}  CX_{a_{ij}b_{i(j+f^{(z)})}} \big) ,
\end{align}
while the single qubit measurements remain the same as in Eq.~\eqref{eq14}.

At this point, we have demonstrated that fractalization relates the trivial history ground state of the 1D cluster state to a nontrivial LCA evolution history ground state of the fractal 2D cluster state. 
The map induced by fractalization in space on the history state, viewed as a process in time, establishes a method to fractalize time that is consistent with the existing method to fractalize spatial diractions~\cite{dom_main}. 

Furthermore, applying fractalization to the symmetries of the 1D cluster state (with open boundary conditions) results in 
\begin{align}
S^{frac}_{1,s} &= \sigma_{Z}[y^{s}\sum_{r}(1+y)^{r}x^{r}\binom{1}{0}]\nonumber\\
S^{frac}_{2,s} &=  \sigma_{X}[y^{s}\sum_{r}(1+y)^{r}x^{r}\binom{0}{1}] .
\end{align}
The fractalized cluster state Hamiltonian $\mathcal{H}_{FCS}$ realises a subsystem symmetry-protected topological (SSPT) phase on the cylinder protected by the set of fractal subsystem symmetries generated by $S^{frac}_{\alpha,s}.$ For $f(w)=1+w$, these symmetries act subsystems that correspond to the discrete Sierpinski fractal, shifted by $s$ along the $y$ direction.

\subsection{Adapative Circuit Implementation}\label{circuitimplementation}
In this section we demonstrate that the time-fractalization of a quantum memory based on a CSS code can be implemented via a constant-depth local adaptive quantum circuit built from products of controlled-$X$ gates over a constant range $r=f^{(n_f)}$ followed by single qubit measurements and a Pauli feed-forward operation. 
The circuit of controlled-$X$ gates is 
\begin{align}
    \prod_{k=0}^{r} C_iX_{i+k}
\end{align}
where $C_iX_{i+k}$ denotes a gate controlled on $i$ with target $i+k$. 
We compose these gates in a ladder fashion to implement the full update rule on a line of qubits indexed by $\mathbb{Z}$ 
\begin{align}
   V =  
   \cdots ( \prod_{z=0}^{n_{f}}  CX_{(j-1)(j-1+f^{(z)})} ) ( \prod_{z=0}^{n_{f}}  CX_{j(j+f^{(z)})} ) \nonumber \\ 
   \times( \prod_{z=0}^{n_{f}}  CX_{(j+1)(j+1+f^{(z)})} )  \cdots
\end{align}

The Sierpinski update rule corresponds to the circuit
\begin{align}
    \includegraphics[page=18,scale=1.5]{Figures.pdf} 
\end{align}
where the qubit index $j$ runs up the page, and time runs from left to right. 
This circuit has linear depth in the number of qubits it is applied to (and there is an ambiguity in the presence of periodic boundary conditions). 
However, it is simple to implement this operator via a constant-depth local unitary circuit followed by measurements and the application of a Pauli-$Z$ feed-forward operator. 
This is based on a simple gadget 
\begin{align}
    \includegraphics[page=19]{Figures.pdf} \label{gadget1}
   \\
   \includegraphics[page=20]{Figures.pdf}
   \label{gadget2}
\end{align}
where the second relation can be used to identify and correct the Pauli-$Z$ byproduct operator that occurs due to the randomness of measurement outcomes (even in the case of error-free measurements). 

Composing the measurement gadgets together we find a constant-depth local unitary circuit on the original qubits plus ancillas, followed by a round of single qubit measurements and a Pauli-$Z$ feedforward operation. This implementation does not require linear depth and can easily be applied with periodic boundary conditions. 
For periodic boundary conditions, the parity of the total number of $\ket{-}$ measurement outcomes cannot be changed by the feed-forward operation. Hence, the quantum channel implemented by the adaptive quantum circuit decomposes into an even and odd component that are supported on the $+1$ and $-1$ charge sectors of a global $\prod X$ symmetry, respectively~\cite{Devakul_2018}. 
For the case of all $\ket{+}$ outcomes, we find the desired operator (up to an overall normalization)
\begin{align}
    \includegraphics[page=21]{Figures.pdf} 
\end{align}
We now relate this to our description of the Sierpinski cluster state from Eqs.~(\ref{eq13},\ref{eq14}) 
\begin{align}
    V_i \ket{\psi}_{a_{i}} \ket{0}^{\otimes n}_{b_{i}}\ket{+}^{\otimes n}_{a_{(i+1)}}:= &\prod_{j} \mathcal{M}{X^+_{a_{ij}}} \mathcal{M}{Z^+_{b_{ij}}} 
    \nonumber \\
    \prod_{j}\big( CX_{a_{ij} b_{ij}} CX_{a_{ij}b_{i(j+1)}} & CX_{a_{(i+1)j} b_{ij}} \big)
    \ket{\psi}_{a_{i}}\ket{0}^{\otimes n}_{b_{i}} \ket{+}^{\otimes n}_{a_{(i+1)}}
    \nonumber \\
    &=V \ket{\psi}. 
\end{align}
That is, by entangling qubits $a_{ij}$  and $a_{(i+1)j}$ to $b_{ij}$, and also entangling $a_{ij}$ to $b_{i(j+1)}$, then measuring the qubits $a_{ij}$ in the $X$ basis, $b_{ij}$ in the Z basis for all sites $ij$ in a single column with fixed $i$, we realise the circuit $V$ (assuming all $+1$ outcomes).

To implement the desired operator in the presence of random measurement outcomes we follow the implementation of $V$ with a feed-forward Pauli operator based on the measurement results. 
This is equivalent to tracking the measurement outcome dependent byproduct operators in an MBQC implementation. 
Each $-1$ measurement result generates a Pauli byproduct operator which can be removed via a Pauli feed-forward operator. Alternatively, these Pauli byproduct operators (the Pauli frame) can be tracked to the final timestep of the evolution such that only a single Pauli feed-forward operator need be applied (equivalently the final measurement basis is adapted). 
This is possible as the evolution of the time-fractalized codes we consider can be described using the stabilizer formalism~\cite{Gottesman1997}. 
We remark that only certain configurations of measurement outcomes can appear, as the measurement outcomes must obey symmetries of the cluster-state if no physical errors are present. 
As noted above, in the above Sierpinski example, the sets of possible measurement outcomes break into two disconnected sectors corresponding to even or odd global parity.

\section{Fractalized Raussendorf-Bravyi-Harrington Cluster State}
\label{concatenation}

In this section we formulate spacetime fractalization of fault-tolerant quantum memories based on CSS codes. As our main example, we consider fractalizing space and time in the (2+1)D toric code quantum memory.  To achieve this spacetime fractalization, we apply space fractalization to the 3D Raussendorf-Bravyi-Harrington (RBH) cluster state and characterize the fractalized Hamiltonian and its symmetries. 
Finally, we describe an adaptive circuit implementation of the spacetime-fractalized toric code quantum memory derived from the fractalized RBH cluster state. 

\subsection{Hamiltonian Description}
\begin{figure}[t]
    \centering
   \includegraphics[page=4,scale=1.5]{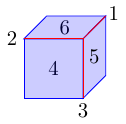} 
    \caption{A labelling of qubits on the edges and faces of a unit cell in the 3D RBH Cluster State lattice.}
    \label{eq15}
\end{figure}

To realise a topologically protected version of the Sierpinski cluster state we consider fractalizing the 3D Raussendorf-Bravyi-Harrington (RBH) cluster state which provides a fault-tolerant, encoded version of the 1D cluster state~\cite{Raussendorf2005,PhysRevLett.86.5188}, see Fig.~\ref{eq15}. We describe the Hamiltonian in CSS form as
\begin{align}
H_{RBH} &= -\sum_{\textbf{r}} \sigma_{X} [x^{\textbf{r}}\textbf{A}(\textbf{x})] +\sigma_{Z} [x^{\textbf{r}}\textbf{B}(\bar{\textbf{x}})]
\end{align}
where
\begin{align}
\textbf{A}(\textbf{x})&=
\begin{bmatrix}
1 & 0 & 0\\
0 & 1 & 0\\
0 & 0 & 1\\
0 & 1+z & 1+y\\
1+z & 0 & 1+x\\
1+y & 1+x & 0
\end{bmatrix}\nonumber\\ 
\textbf{B}(\bar{\textbf{x}})&=
\begin{bmatrix}
    0 & 1+\bar{z} & 1+\bar{y}\\
    1+\bar{z} & 0 & 1+\bar{x}\\
    1+\bar{y} & 1+\bar{x} & 0 \\1 & 0 & 0\\
    0 & 1 & 0\\
    0 & 0 & 1
\end{bmatrix}.
\end{align}

Computation with the cluster state proceeds by measuring each qubit in either the $X$ or $Z$ basis. 
The qubits corresponding to the first three rows of the stabilizer matrices are measured in the $Z$ basis, and the qubits corresponding to the latter three rows are measured in the $X$ basis. 
These single qubit measurements propagate the logical information through the cluster state, driving the computation~\cite{PhysRevLett.86.5188}.
The local 1-form symmetries of the state are generated by 
\begin{align}
&X[L_{1}(\textbf{x})]=\sigma_{X}\begin{bmatrix}
    1+x\\
    1+y\\
    1+z\\
    0\\
    0\\
    0
\end{bmatrix},
&Z[L_{1}(\bar{\textbf{x}})]=\sigma_{Z}\begin{bmatrix}
    0\\
    0\\
    0\\
    1+\bar{x}\\
    1+\bar{y}\\
    1+\bar{z}
\end{bmatrix},
\end{align}
and representative global symmetries are
\begin{align}
&X[S_{1}(\textbf{x})]=\sigma_{X} \begin{bmatrix}
    \sum_{i,j}y^{i}z^{j} & 0 & 0 \\
    0 & \sum_{i,j}x^{i}z^{j} & 0\\
    0 & 0 & \sum_{i,j}x^{i}y^{j}\\
    0 & 0 & 0 \\
    0 & 0 & 0\\
    0 & 0 & 0
\end{bmatrix},\nonumber\\
&Z[S_{2}(\textbf{x})]=\sigma_{Z} \begin{bmatrix}
    0 & 0 & 0\\
    0 & 0 & 0\\
    0 & 0 & 0\\
    \sum_{i,j}y^{i}z^{j}&0&0\\
    0 & \sum_{i,j}x^{i}z^{j} & 0\\
    0 & 0 & \sum_{i,j}x^{i}y^{j}
\end{bmatrix} \label{logicalsnormal}
\end{align}
see Ref.~\cite{Roberts_2017,Roberts_2020}.

Fractalizing the RBH cluster state produces the Hamiltonian
\begin{align}\label{eq19}
H_{RBH}^{frac} &= -\sum_{ij} \sigma_{X} [x^{\textbf{r}}w^{s}\textbf{A}(\textbf{f}(w)\circ \textbf{x})] \nonumber \\ &\qquad\qquad\quad +\sigma_{Z}[x^{\textbf{r}}w^{s}\textbf{B}({\textbf{f}}(\bar{w})\circ \bar{\textbf{x}})] 
\end{align}
where 
\begin{align}
    \textbf{f}(w)\circ \textbf{x} = 
    \begin{bmatrix}
        f(w) \circ x\\
        g(w) \circ y\\
        h(w) \circ z
    \end{bmatrix}.
\end{align}
The fractalized RBH model inherits fractalized version of the local and global symmetries 
\begin{align}\label{fracsym}
    &X[L_{1}(\textbf{f}(w)\circ\textbf{x})],Z[L_{2}({\textbf{f}}(\bar{w})\circ\bar{\textbf{x}})],\nonumber\\
    &X[S_{1}(\textbf{f}(w)\circ\textbf{x})],Z[S_{2}({\textbf{f}}(\bar{w})\circ\bar{\textbf{x}})].
\end{align}

The fractalized RBH cluster state specifies a dynamical quantum process via MBQC. MBQC proceeds by measuring each qubit of the cluster state in either the $X$ or $Z$ basis, as described above, thus driving the computation by teleporting the logical state encoded on the initial timeslice to the final timeslice. However, the MBQC implementation relies on an additional layer of complexity since the arbitrary measurement outcomes on each measured qubit affect the Pauli frame of the logical state at the final timeslice. The Pauli frame can be reconstructed by keeping track of all measurement outcomes. This can be done for the fractalized Raussendorf cluster state as well as the non-fractalized cluster state as shown in Appendix~\ref{tensornetwork}. 

\subsection{Adaptive Circuit Realisation}\label{circuittoriccode}

\begin{figure}[t]
    \centering
    \includegraphics[page=14,scale=1.5]{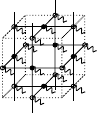} 
    \caption{Unit cell of the Raussendorf cluster state tensor network. Primal qubits are filled nodes and dual qubits are empty nodes. Solid edges denote virtual indices and wavy edges denote physical indices. }
    \label{fig:raussunitcell}
\end{figure}

We require an adaptive Floquet circuit in order to fractalize the (2+1)D spacetime process specified by the Raussendorf cluster state and implement it via a constant depth, adaptive local circuit. In Section~\ref{circuitimplementation} we described how to explicitly implement fractalization on the one dimensional cluster state. By applying tensor network identities, we can modify the fractalized RBH cluster state to obtain the following circuit procedure: First measure the fractalized $X$ stabilizers of the surface code with respect to the $x$ and $y$ variables. Next, measure the fractalized $Z$ stabilisers of the surface code with respect to the $x$ and $y$ variables. Finally, apply the fractalized time evolution in the $z$ variable using transversal CNOT and measurement operations, and repeat. We use the CNOT gadget from Eq.~\ref{gadget1} to realise the transversal CNOTs and measurements in a local, constant depth manner. Below, we use the convention that $\bullet$ nodes of the tensor network represent qubits initialised in the logical $|+\rangle$ state, and $\oplus$ nodes represent qubits initialised in the logical $|0\rangle$ state. We use the following $n$-index tensors
\begin{align}
\vcenter{\hbox{\includegraphics[page=16]{Figures.pdf} }}  &=\prod_{i=1}^{n}\delta(a_{i}+a_{i+1}),
\\
\vcenter{\hbox{\includegraphics[page=17]{Figures.pdf} }} &=\delta\left(\sum_{i=1}^{n}a_{i}\right).
\end{align}

We now consider an example specified by ${f(w) = g(w) = 1,}$ $h(w) = 1+ w$. 
In this example only the time dimension is fractalized nontrivially. 
This produces a floquet code corresponding to layers of the surface code undergoing a fractalized time evolution. 

\begin{figure}[t]
    \centering
    \includegraphics[page=12,scale=1.5]{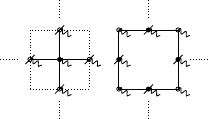} 
    \caption{Slices of a unit cell of the lattice at $t=0$ (left) and $t=1$ (right). The unit cell slices are the same for $w=0,1,2.$}
    \label{fig:fig8}
    \end{figure}

\begin{figure}[t]
    \includegraphics[page=13,scale=1.5]{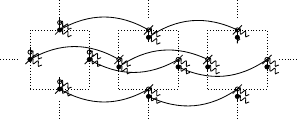} 
    \caption{Slices of a unit cell of the lattice at $t=1.5$ for $w=0$ (left) $w=1$ (center) and $w=2$ (right).}
    \label{fig:fig9}
\end{figure}

A naive approach to fractalizing the cluster state applies the CNOT gadget in Eq.~\ref{gadget1} between qubits that are entangled by the LCA update rule. However, this entangles qubits across different time steps. We now modify this approach to obtain fractalization circuits that consist wholly of CNOTs acting on qubits at equal timesteps. To accomplish this we apply tensor identities to split each qubit at the odd timesteps of the cluster state into two as shown in Figs.~\ref{fig:fig8},~\ref{fig:fig9}, such that we have two rounds of stabiliser measurement at $t = 0 ,1$ followed by a fractalization step at $t = 1.5,$ followed by two more rounds of stabiliser measurement, and so on. In Figs.~\ref{fig:fig8},~\ref{fig:fig9}, the dashed lines represent the structure of the lattice while the bold lines represent the physical CNOT gates.

This procedure can be naturally generalised to include arbitrary functions $f(w),g(w),h(w).$ In the case where ${f(w)=g(w) = 1}$ and $h(w)$ is nontrivial, Fig.~\ref{fig:fig8} is unchanged and implements stabiliser measurements of the cluster state. However, Fig.~\ref{fig:fig9} is modified to implement the fractalizing CNOTs specified by $h(w).$ 
Similarly, when $f(w),g(w),$ are nontrivial and $h(w)=1$, Fig.~\ref{fig:fig8} is modified to implement measurement of the Fractal Spin Liquid checks specified by $f(w)$ and $g(w)$~\cite{Yoshida_2013}. In this case, the entanglement between layers depicted in Fig.~\ref{fig:fig9} remains unchanged. 

When $f(w),g(w),$ and $h(w)$, are all nontrivial, the timesteps cannot be partitioned into `non-fractalizing' and `fractalizing' steps. Specifically, Fig.~\ref{fig:fig8} now includes additional CNOTs across additional lattice sites on the same timeslice that realise the fractalizing $f(w),g(w),$ checks of the LCA. Therefore measuring individual qubits of the cluster state at $t=0,1,$ effectively measures the stabilizers that have been fractalized in the space direction. Measuring the cluster state at $t=1.5$ fractalizes the cluster state in time.

\begin{figure}[t]
    \centering
    \includegraphics[page=29,scale=1]{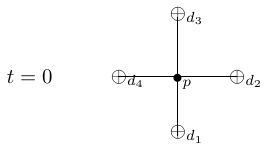} 
    \caption{Slices of the unit cell at even (E) times. Primal and dual qubits labelled.}
    \label{fig:even}
\end{figure}

\begin{figure}[t]
    \centering
    \includegraphics[page=30]{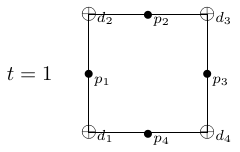} 
    \caption{Slices of the unit cell at odd (O) times. Primal and dual qubits labelled.}
    \label{fig:odd}
\end{figure}

We now describe the implementation of a single floquet period of the fractalized RBH time evolution using constant depth adaptive circuits, following Sec.~\ref{Vformalism}. First, we introduce notation to index the positions of qubits in the RBH cluster state. Consider a unit cell of the cluster state, as depicted in Fig.~\ref{fig:raussunitcell}. Call the timeslice at $t=0$ the even timeslice, denoted by $E$, and the timeslice at $t=1$ the odd timeslice, denoted by $O$. The $E$ slice has a single primal qubit and four dual qubits, labelled as in Fig.~\ref{fig:even}. The $O$ time slice has four primal and four dual qubits, labelled as in Fig.~\ref{fig:odd}. The entangling gates that initialize each timeslice of the cluster state are 
\begin{align}
    &\mathcal{I}=\prod_{ijkw}\left(\prod_{a=1}^{4}CX_{E^{(ijkw)}_{p}E^{(ijkw)}_{d_a}}\right)\\\nonumber
    &\times\left(\prod_{a=1}^{4}CX_{0^{(ijkw)}_{p_{a}}0^{(ijkw)}_{d_{a}}}CX_{0^{(ijkw)}_{p_{a}}0^{(ijkw)}_{d_{a+1}}}\right)\\\nonumber
    &=:\prod_{ijkw}I_{O}^{(ijkw)}I_{E}^{(ijkw)},
\end{align}
where the operators $I_{O}^{(ijkw)},I_{E}^{(ijkw)}$ represent entangling the $O,E$ faces of a unit cell located at $ijkw$ in space at a fixed timeslice. 
We then express the product of operators representing the application of entangling gates between adjacent timeslices as
\begin{align}
&\mathcal{U} =\prod_{ijkw}\left(\prod_{a=1}^{4}CX_{E^{(ijkw)}_{d_a}O^{(ijkw)}_{p_a}}\right)\\\nonumber
&\times\left(\prod_{a=1}^{4}CX_{O^{(ijkw)}_{p_a}E^{(ij(k+1)w)}_{d_a}}\right)\\\nonumber
&=\prod_{ijkw}U_{EO}^{(ijkw)}U_{OE}^{(ijkw)}
\end{align}
where $U_{EO}^{(ijkw)},U_{OE}^{(ijkw)}$ represent entangling timeslices of a unit cell from $E$ to $O$ and $O$ to $E$ timeslices. The final component we require are the entangling gates specified by the fractalizing adaptive local circuit, which for fractalization in time is
\begin{align}
    &\mathcal{F}_{h}=\prod_{ijkw}\left(\prod_{b=1}^{n_h}\prod_{a=1}^{4}CX_{O^{(i,j,k,w)}_{p_a}E^{(i+1,j,k,h^{(b)}(w))}_{d_a}}\right)\\\nonumber
    &=\prod_{ijkw}F^{(i,j,k,w)}_{OE,h(w)}.
\end{align}
For fractalization in the $x$ direction for $O$ and $E$ timeslices respectively we have
\begin{align}&F_{EE,f(w)}^{(i,j,k,w)}=\prod_{b=1}^{n_{f}}\prod_{a=1}^{4}CX_{E^{(i,j,k,w)}_{p}E^{(i+1,j,k,f^{(b)}(w)}_{p}}\\\nonumber
    &\times CX_{E^{(i,j,k,w)}_{d_a}E_{d_a}^{(i+1,j,k,f^{(b)}(w)}}\\\nonumber
&F_{OO,f(w)}^{(i,j,k,w)}=\prod_{b=1}^{n_{f}}\prod_{a=1}^{4}CX_{O^{(i,j,k,w)}_{p_a}O^{(i+1,j,k,f^{(b)}(w)}_{p_a}}\\\nonumber
    &\times CX_{O^{(i,j,k,w)}_{d_a}O_{d_a}^{(i+1,j,k,f^{(b)}(w)}}\\
\end{align}
such that
\begin{align}\mathcal{F}_{f}&=\prod_{ijkw}F^{(i,j,k,w)}_{EE,f(w)}F^{(i,j,k,w)}_{OO,f(w)}
\end{align}
and similarly for fractalization in the $y$ direction
\begin{align}
&F_{EE,g(w)}^{(i,j,k,w)}=\prod_{b=1}^{n_{f}}\prod_{a=1}^{4}CX_{E^{(i,j,k,w)}_{p}E^{(i,j+1,k,g^{(b)}(w)}_{p}}\\\nonumber
    &\times CX_{E^{(i,j,k,w)}_{d_a}E_{d_a}^{(i,j+1,k,g^{(b)}(w)}}\\\nonumber
&F_{OO,g(w)}^{(i,j,k,w)}=\prod_{b=1}^{n_{f}}\prod_{a=1}^{4}CX_{O^{(i,j,k,w)}_{p_a}O^{(i,j+1,k,g^{(b)}(w)}_{p_a}}\\\nonumber
    &\times CX_{O^{(i,j,k,w)}_{d_a}O_{d_a}^{(i,j+1,k,g^{(b)}(w)}}\\\nonumber
\mathcal{F}_{g}&=\prod_{ijkw}F^{(i,j,k,w)}_{EE,g(w)}F^{(i,j,k,w)}_{OO,g(w)}.
\end{align}
Given these operators we can express the full set of entangling gates as
\begin{align}  \mathcal{E}_{fgh}&=(\mathcal{F}_{f}\mathcal{F}_{g}\mathcal{F}_{h})\cdot\mathcal{U}\cdot\mathcal{I}.
\end{align}
We now describe the set of measurements. Given the entangled cluster state, computation proceeds by measuring primal qubits in the $X$ basis and dual qubits in the $Z$ basis. Explicitly,
\begin{align}
&\mathcal{M}=\prod_{ijkw}\prod_{a=1}^{4}M_{X_{E_p}^{(ijkw)}}M_{Z_{E_{d_a}}^{(ijkw)}}M_{X_{O_{p_a}}^{(ijkw)}}M_{Z_{O_{d_a}}^{(ijkw)}}\\\nonumber
&=\prod_{k}\mathcal{M}{E}^{(k)}\mathcal{M}{O}^{(k)}
\end{align}
where we have defined
\begin{align}
    \mathcal{M}{E}^{(i,j,k,w)}&=\prod_{a=1}^{4}M_{X_{E_p}^{(ijkw)}}M_{Z_{E_{d_a}}^{(ijkw)}}\\\nonumber
    \mathcal{M}{O}^{(i,j,k,w)}&=\prod_{a=1}^{4}M_{X_{O_{p_a}}^{(ijkw)}}M_{Z_{O_{d_a}}^{(ijkw)}}.
\end{align}
Then for a cluster state unit cell with timeslices initialised by $\mathcal{I},$ applying $U_{EO}^{(ijkw)},F_{EE,f(w)}^{(i,j,k,w)},F_{EE,g(w)}^{(i,j,k,w)}$ and measuring $\mathcal{M}{E}^{(i,j,k,w)}$ both teleports the logical state from the $E$ to $O$ timeslice and realises fractalization in the $x,y$ directions. Similarly, applying $U_{OE}^{(ijkw)},F_{OE,h(w)}^{(i,j,k,w)},F_{OO,f(w)}^{(i,j,k,w)},F_{OO,g(w)}^{(i,j,k,w)}$ teleports the logical state from the $E$ to $O$ timeslice and realises fractalization in the $x,y$ and time directions.

We have not yet discussed byproduct operators due to the random measurement outcomes for clarity. It is straightforward to deal with the random measurement outcomes by keeping track of measurement outcomes on the ancilla qubits required to implement the CNOT gadgets, and commuting certain measurement outcome dependent byproduct operators through time to the final timeslice, as described in Appendix~\ref{tensornetwork}.

\section{Excitations and Logical Operators}
\label{sec:Excitations}

In this section, we describe how fractalization affects the logical operators and excitations of the 3D Raussendorf cluster state. We first construct an operator that keeps track of logical operator evolution in both the regular and fractalized codes.
We then describe the evolution of excitations in both the regular and fractalized codes. We show that by fractalizing the 3D Raussendorf cluster state, subject to certain independence conditions, one obtains a code with fully immobile fracton excitations in space and time leading to a superlinear spacetime fault-distance. 

\subsection{Logical Operators}\label{logicals}
\begin{figure}[t]
    \centering
    \includegraphics[page=5,scale=1.8]{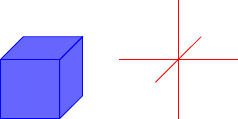} 
    \caption{Z type local symmetry (left) and X type local symmetry (right).}
    \label{fig:11}
\end{figure}

Consider a fractalized cluster state that is infinite in the $x$ and $y$ directions, and finite in time such that $x,y\in(-\infty,\infty), t\in[0,m].$ Suppose that the boundaries at $t=0,m,$ are smooth. Let $\vec{e}_{i}$ denote a 6-entry column vector with a $1$ in the $ith$ column and zeros elsewhere. The local symmetries of the RBH cluster state are shown in Fig.~\ref{fig:11}. Whilst the full cluster state does not encode any logical qubits, we can think of the $t = 0,m$ boundaries encoding $k$ maximally entangled qubits. The logical operators for these encoded qubits are given by the restriction of the global symmetries to the boundaries. In particular, consider the restriction of the global symmetries 
\begin{align}
X_{1} &= \sigma_{X}
    \left[
       \sum_{ij}g^{i}(w)y^{i}h^{j}(w)z^{j}\big(\vec{e}_{1}\big)\right]\\
    X_{2} &= \sigma_{X}\left[
       \sum_{ij}f^{i}(w)x^{i}h^{j}(w)z^{j}(\vec{e}_{2})\right]\\
     Z_{2} &= \sigma_{Z}\left[
       \sum_{ij}g^{i}(w)y^{i}h^{j}(w)z^{j}(\vec{e}_{4})\right]\\
      Z_{1} &= \sigma_{Z}\left[
       \sum_{ij}f^{i}(w)x^{i}h^{j}(w)z^{j}(\vec{e}_{5})\right]
\end{align}
to the boundaries $t = 0,m.$ For the non-fractalized case, there are two pairs of logical operators which are exactly those of the toric code. 
When the cluster state is fractalized only in time, ie $f(w) = g(w) = 1$ and $h(w)$ non-trivial, since we are restricting to $z = 0$ the boundary operators introduced above apply the $t = 0$ boundary, for each value of $w$. However, their partner logical operators on the $t = m$ boundary evolve a more complex structure due to the fractalization in time. Taking $h(w) = 1+w$ gives $h^{j}(w)z^{j} = (1+w)^{m}z^{m}$ so in addition to the boundary terms at $w = k$, each boundary operator has support on qubits with the same $(x,y,z)$ locations, at $w = k +m.$

We now describe how to determine whether a given logical has been flipped at the final timeslice. This is done by keeping track of the global symmetries of the cluster state as qubits are measured. Consider first the non-fractalized case. We take the symmetry
\begin{align}
    \sigma_{X}\left[\sum_{ij}y^{i}z^{j}(\vec{e}_{1})\right]
\end{align}
as an example. This symmetry tells us that for a fixed $x$ value, say $x = k$, the product of measurement outcomes of the specified qubits on the plane up to time $t$ must be equal to the eigenvalue of the product of $X$s on the plane after the time $t.$ Let the product of all $X$s measured on qubits in the plane $\sum_{ij}y^{i}z^{j}(\vec{e}_{1})$ up to time $t$ be denoted by
\begin{align}
    S_{X}^{1,t,k} = \sigma_{X}\left[x^{k}\sum_{i=0}^{\infty}\sum_{j=0}^{t}y^{i}z^{j}(\vec{e}_{1})\right].
\end{align}
Let the eigenvalue of the product of $X$s on the plane after time $t$ be denoted by
\begin{align}
    \bar{S}_{X}^{1,\bar{t},k} = \sigma_{X}\left[x^{k}\sum_{i=0}^{\infty}\sum_{j=t}^{m}y^{i}z^{j}(\vec{e}_{1})\right].
\end{align}
Then we have $S_{X}^{1,t,k}\cdot \bar{S}_{X}^{1,\bar{t},k}=\mathbb{I}$ for all $k,t.$ In particular, if $S_{X}^{1,m-1,k} = -\mathbb{I},$ the corresponding logical operator needs to be flipped. Analogous statements are true for the other logicals.

The extension to the fractalized  case is straightforward. The symmetries are given by Eq.~\ref{fracsym}. We can write
\begin{align}
    S_{X}^{1,t,k} = \sigma_{X}\left[x^{k}\sum_{i=0}^{\infty}\sum_{j=0}^{t}y^{i}g(w)^{i}z^{j}h(w)^{j}(\vec{e}_{1})\right]\\
    \bar{S}_{X}^{1,\bar{t},k} = \sigma_{X}\left[x^{k}\sum_{i=0}^{\infty}\sum_{j=t}^{m}y^{i}g(\bar{w})^{i}z^{j}h(\bar{w})^{j}(\vec{e}_{1})\right],
\end{align}
where we have $S_{X}^{1,t,k}\cdot \bar{S}_{X}^{1,\bar{t},k}=\mathbb{I}$. Consider the specific example $h(w) = 1+w, g(w) = f(w) = 1.$ Then the symmetry $S_{X}^{1,t,k}$ acts in the $w$ dimension on a subset of sites corresponding to the Sierpinski fractal. The logical corresponding to $S_{X}^{1,t,k}$ needs to be flipped at $t=m$ if
\begin{align}
    S_{X}^{1,m-1,k} = \sigma_{X}\left[x^{k}\sum_{i=0}^{\infty}\sum_{j=0}^{m-1} y^{i}z^{j}(1+w)^{j}(\vec{e}_{1})\right] = -\mathbb{I}.
\end{align}
Analogous conditions hold for the other logical operators.

\subsection{Excitations}
In the following, we describe how fractalization affects the evolution of excitations in the cluster state. We follow the approach of Ref.~\cite{Yoshida_2013} and adapt it to include a time dimension. We begin by detailing the non-fractalized case, before extending the description to the fractalized case, for which we find excitations that are immobile in both space and time. 
\subsubsection{RBH Cluster State}
\begin{figure}[t]
    \centering
    \includegraphics[page=6,scale=1.8]{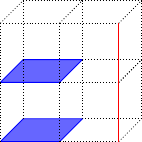} 
    \caption{Z excitation (left) propagating from $\bar{z}\vec{e}_{6}$ to $\vec{e}_{6}.$ X excitation (right) propagating from $\vec{e}_{3}$ to $z\vec{e}_{3}.$ }
    \label{fig:12}
\end{figure}

The evolution of excitations in time in the cluster state is determined by the local symmetries.
Consider an arbitrary $Z$ error $Z(e_{A},e_{B},e_{C},e_{D},e_{E},e_{F})^{T}.$ The excitation polynomial is given by the convolution of the error with the $Z$ type local symmetry matrix. Specifically, we have
\begin{align}
        &E(x,y,z,w) = 
    \begin{bmatrix}
        e_{A}\\
        e_{B}\\
        e_{C}\\
        e_{D}\\
        e_{E}\\
        e_{F}
    \end{bmatrix}^{T}\cdot
 \begin{bmatrix}
0\\
0\\
0\\
1+x\\
1+y\\
1+z
\end{bmatrix}\nonumber\\
&=e_{D}(1+x)+
    e_{E}(1+y)+
    e_{F}(1+z).
\end{align}
Suppose that we have an error $X_{e} = \sigma_{X}[\vec{e}_{1}]$ at $(x,y,z) = (0,0,0).$ Then the excitation polynomial will be $E(x,y,z,w) = 1+z.$ This means that if there is an isolated excitation at $(0,0,0),$ represented by $1$, the application of $X_{e}$ causes it to propagate to an excitation at $(0,0,1),$ represented by $z$.  In particular, excitations are pointlike and can be moved by stringlike faults. In the MBQC picture, an $X_e$ error on an edge in the time direction is equivalent to a measurement fault in a circuit implementation. The situation is analogous for local $X$ symmetries and $Z$ errors, see Fig.~\ref{fig:12}.

\subsubsection{Fractalized RBH Cluster State}

\begin{figure}[t]
    \centering
    \includegraphics[page=7,scale=1.8]{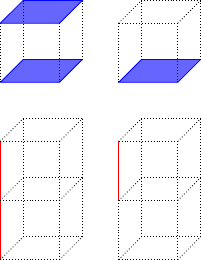} 
    \caption{Z excitation (top) propagating from $\bar{z}\vec{e}_{6},\bar{z}\bar{h}(w)\vec{e}_{6}$ to $\vec{e}_{6}.$ X excitation (bottom) propagating from $\vec{e}_{3}$ to $z\vec{e}_{3},zh(w)\vec{e}_{3}.$ }
    \label{fig:13}
\end{figure}

We now consider the fractalized RBH model. The excitation polynomial becomes
\begin{align}
        &E(x,y,z,w) = 
    \begin{bmatrix}
        e_{A}\\
        e_{B}\\
        e_{C}\\
        e_{D}\\
        e_{E}\\
        e_{F}
    \end{bmatrix}^{T}\cdot
 \begin{bmatrix}
0\\
0\\
0\\
1+f(w)x\\
1+g(w)y\\
1+h(w)z
\end{bmatrix}\nonumber\\
&=e_{D}(1+f(w)x)+
    e_{E}(1+g(w)y)+
    e_{F}(1+h(w)z).
\end{align}
We consider an error $X_{e} = \sigma_{X}\left[\vec{e}_{1}\right]$. The excitation polynomial is $E(x,y,z,w) = 1+h(w)z.$ Under the application of $X_e$, an isolated pointlike excitation at $(0,0,0,0)$, described by 1, propagates to multiple excitations represented by $h(w)z$, see Fig.~\ref{fig:13}. The excitation cannot be moved forward in time without creating additional excitations, which incurs an additional energy cost. An isolated excitation is therefore immobile. Similar to the paragraph above, in a circuit implementation an $X_e$ error in the time direction becomes equivalent to a measurement fault. The case is analogous for local $X$ symmetries and $Z$ faults. Hence, in a nontrivially fractalized RBH model isolated syndromes are immobile in the time direction. Specifically for an isolated excitation, $f(w)$ is applied when propagating in the $x$ direction, $g(w)$ is applied when propagating in the $y$ direction and $h(w)$ when propagating in the $z$ direction. If $f(w),g(w),h(w)$ are algebraically independent, excitations in the fractalized model are immobile in both space and time~\cite{Yoshida_2013}. In this case we say the model is intrinsically \textit{spacetime~Type-II}. This means there is no string-like logical fault in the spacetime model. 

\begin{figure}[t]
    \centering
    \raisebox{1.15cm}{\includegraphics[page=8,scale=1.3]{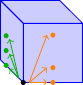}}
    \ 
    \includegraphics[page=9,scale=1.3]{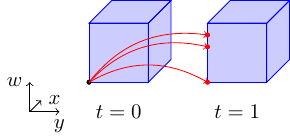} 
    \caption{Hopping dynamics of a single excitation in space (left) in the $\hat{x}-\hat{w},\hat{y}-\hat{w}$ directions and in time (right) in the $\hat{z}-\hat{w}$ direction under the application of algebraically independent polynomials $f(w),g(w),h(w).$ Due to the independence of $f(w),g(w),h(w),$ the evolved fracton can never be brought back to a single excitation by the application of any local operator.}
    \label{fig:hopping}
\end{figure}

We now consider an example describing the evolution of a 1D excitation $e(w),$ located at $j = l =k= 0.$ The excitation polynomial is
\begin{align}
    E(x,y,z,w) = e(w)f^{j'}(w)g^{l'}(w)h^{k'}(w)x^{j'}y^{l'}z^{k'}.
\end{align}
We assume that $f(w),g(w),h(w),$ are positioned at the origin such that they have only nonzero constant terms and positive powers. Polynomials are said to be algebraically related if
\begin{align}
    f(w)^{n_{1}} = cg(w)^{n_{2}}h(w)^{n_{3}}
\end{align}
for some finite $n_{i},c$ where $n_{i}\in\mathbb{Z}^{+}$ and not all $n_{i} = 0,$ or for some permutation of $1,2,3$ without periodic boundary conditions. 
An explicit example of three polynomials that are algebraically independent on periodic boundary conditions $f(w)^{L_{i}}=g(w)^{L_{i}}=h(w)^{L_{i}}=1$ where $L_{i}=2^{l}$ is
\begin{align}\label{independentpolys}
    f(w) &=1+w+w^{2}\nonumber\\
    g(w)&=1+w+w^{3}\nonumber\\
    h(w)&=1+w^{2}+w^{3}.
\end{align}
If $f(w),g(w),h(w),$ are algebraically related, the excitation can propagate freely in the $n_{1}\hat{x}-n_{2}\hat{y}-n_{3}\hat{z}$ direction. If $f(w),g(w),h(w),$ are not algebraically related, the excitation cannot propagate in any direction without creating additional excitations. This is depicted in Fig.~\ref{fig:hopping}, which adopts an adaptive local circuit of the form in Eq.~\ref{independentpolys}. On the left of Fig.~\ref{fig:hopping}, the cascading of a single excitation in space in the $\hat{x},\hat{y}$ directions due to application of $f(w),g(w)$ is shown. On the right of Fig.~\ref{fig:hopping}, we see the cascading of a single excitation in the $\hat{z}$ time direction under the application of $h(w)$. The three polynomials chosen above are algebraically independent, so evolved fracton can never be brought back to a single excitation by the application of any local operator.
Therefore, the model is intrinsically spaceimte Type-II, which is equivalent to the absence of any string-like logical operator segments in space and time.

\section{Discussion and Conclusion}
\label{sec:disc} 

In this work we have extended fractalization from space to spacetime~\cite{dom_main,Yoshida_2013}. We focused on applying fractalization to the 3D RBH cluster state for which we gave Hamiltonian and a tensor network descriptions, analysed the evolution of excitations and logical operators, and proposed an adaptive local unitary circuit implementation. This led to the construction of models with generalised fracton excitations that are immobile in both space and time. The fractal hopping patterns of fracton excitations in time lead to spacetime topological phases of matter with response periods which scale exponentially in system size~\cite{LCA}. This extends the notion of a discrete time crystal~\cite{dct1,dct2,dct3}, a periodically driven Floquet system with a stable steady state that has a longer period than the drive period, which breaks discrete time translational symmetry. Discrete time crystals are characterised by a response period which is an integer multiple of the drive period~\cite{Else_2020}. 

Here, we have introduced spacetime fracton models that can be implemented via adaptive local unitary circuits which generalize a floquet drive period. These models exhibit an extreme form of discrete time translation symmetry breaking, with a response period that can be exponentially larger than the drive period as a function of system size. Notably, in the continuum limit where the spatial lattice spacing and fundamental period time are taken to zero with the system size $L$ held constant, the spacetime fracton phase breaks a continuous time translation symmetry down to a response period that is a function of $L$. However, we note in this case the continuous time evolution may not correspond to evolution under a physical periodic Hamiltonian.

Fractalization in time is also relevant for the fault tolerant properties of quantum error correction procedures in space and time. In particular, spacetime type-II fracton codes are promising candidates for fault-tolerant quantum information storage. This is due to their superlinear timelike fault-distance scaling which has the potential to lower the time overhead required to achieve a target fault distance. 

Future directions include a thorough fault tolerance analysis of fractalized quantum codes including the fractalized 3D RBH model~\cite{Hillmann_2025}, an investigation of the self-correcting properties of spacetime fractalized quantum memories, the introduction of boundaries in spacetime~\cite{Bulmash_2019,Aitchison_2024}, and the extension of bifurcating entanglement renormalization to spacetime lattices~\cite{Haah_2014,Dua_2020,San_Miguel_2021}. In addition, the generalization of the spacetime fractalization framework to non CSS codes remains open. Finding such a generalization could lead to the discovery of further exotic spacetime topological phases of matter.

\acknowledgments
While this work was in progress some related work on improving timelike distance appeared~\cite{xu2026frameworklowoverheadquantumfault}.
DJW is supported by the Australian Research Council Discovery Early Career Research Award (DE220100625).

\bibliography{bibliography}

\appendix
\onecolumngrid
\section{Tensor Network Description}\label{tensornetwork}
Here, we present a tensor network formulation of the fractalized 3D Raussendorf cluster state. The aim of this formulation is to describe the full cluster state with general measurement outcomes. We describe how to commute the measurement byproduct operators to the final timeslice such that the output state is equivalent to the one produced by a cluster state with ideal measurement outcomes. This is achieved by describing the Pauli byproduct operators on ancilla cluster qubits, and an operator that keeps track of the Pauli frame at the final timeslice. We develop this description on the regular RBH lattice before extending it to the fractalized model.

\subsection{Regular 3D Raussendorf Model}
\begin{figure}[H]
    \centering
    \includegraphics[page=14,scale=2]{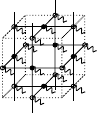} 
    \caption{Unit cell of the 3D Raussendorf cluster state lattice. The graphical conventions follow those in Section~\ref{concatenation}.}
    \label{fig:raussunitcell2}
\end{figure}

We examine a model which is infinite in the x and y directions, and finite in time. Specifically, $x,y\in(-\infty,\infty),t\in[0,m].$ The checks of the model are as per Eq. \ref{eq15}.
A unit cell of the RBH cluster state (without measurements) can be represented as per Fig.~\ref{fig:raussunitcell2}.
The tensor network diagram for the $t=0$ and $t=1$ slices of the unit cell, with measurement outcomes, are
\begin{align}
    \centering
    \includegraphics[page=22]{Figures.pdf} \qquad
    \includegraphics[page=23]{Figures.pdf} .
\end{align}
Above, we have defined the tensors
\begin{align}
    \includegraphics[page=24]{Figures.pdf} \qquad
    \includegraphics[page=25]{Figures.pdf} 
\end{align}
where the left figure depicts a contraction of tensors representing measurement of $|0\rangle$ state with a general measurement outcome determined by $X^{\mu_{P^{i,j,k}_{d_{l}}}}.$ Here, $\mu\in\{\beta,\gamma\}$ and $P\in\{O,E\}$, $l\in\{1,2,3,4\}$. Similarly, the right figure above depicts a contraction of tensors representing measurement of $|+\rangle$ state with a general measurement outcome determined by $Z^{\nu_{P^{i,j,k}_{p_{l}}}}.$ Here, $\nu\in\{\alpha,\delta\}$ and $P\in\{O,E\}$, $l\in\{1,2,3,4\}$. 
Above, we use the same qubit indexing as described in Sec. \ref{circuittoriccode}. Each cell of the lattice has an $(i,j,k)$ index denoting its $x,y,z$ coordinates. It is then assigned an $O,E$ label depending on whether it is an odd or even timeslice. Finally, it is assigned a primal label $p$ or dual label $l$ with possible indexing such that each qubit is indexed in the following way $P^{i,j,k}_{p_{l}/d_{l}}$ where $P=O,E$ and $l=1,2,3,4.$
The powers of X and Z appearing before the measurements indicate the possible effect of a single qubit error flipping the measurement outcome. On even layers of the lattice, powers are $\alpha$ for primal qubits, and $\beta$ for dual qubits. On odd layers of the lattice, powers are $\gamma$ for dual qubits and $\delta$ for primal qubits. We denote $\vec{\alpha} = \{\alpha_{P^{i,j,k}_{p_{l}/d_{l}}}\}_{i,j\in(-\infty,\infty),k\in[0,m],P\in\{O,E\},l\in\{1,2,3,4\}.}$ and similarly $\vec{\beta},\vec{\gamma},\vec{\delta}.$

We aim to commute the $X^{\beta},Z^{\delta}$ through the cluster state onto the $t = m$ slice. This results in a product of Pauli operators acting on the $t = m$ lattice slice, as well as an accumulation of $X^{\beta}$ acting on dual qubits in odd slices. We also keep track of the $Z^{\alpha}$ operators on primal qubits at even time slices, and $X^{\gamma}$ operators on primal qubits on odd time slices.

The operator keeping track of the measurement outcomes $Z^{\alpha},X^{\gamma}$ is
\begin{align}
    \sigma(\vec{\alpha},\vec{\gamma})&=\prod_{l=1}^{4}\prod_{i,j=-\infty}^{\infty}\prod_{k=0}^{m}Z_{E^{i,j,k}_{p}}^{\alpha_{E^{i,j,k}_{p}}}X_{O^{i,j,k}_{p_{l}}}^{\gamma_{O^{i,j,k}_{p_{l}}}}.
\end{align}
For each location of a dual qubit on an even face we get an operator that keeps track of the commutators on dual qubits resulting from propagating the $X^{\beta},Z^{\delta}$ through to $t = m.$
    This operator is given by
\begin{align}
    \omega_{i,j,l}(\vec{\beta},\vec{\delta}) = \prod_{k=0}^{m-1}(X^{\beta_{E^{i,j,k}_{d_{l}}}})^{\otimes {E^{i,j,k'}_{d_{l}}}:k'>k}
    (X^{\beta_{E^{i,j,k}_{d_{l}}}})^{\otimes O^{i,j,k'}_{d_{r}}: O^{i,j,k}_{d_{r}}\in\mathcal{N}(E^{i,j,k}_{d_{l}}),k'>k}
    (Z^{\delta_{O^{i,j,k}_{p_{r}}}})^{\otimes {O^{i,j,k'}_{p_{r}}}:k'>k}.
\end{align}

The final operator needed to describe the state is the product of Paulis commuted to $t = m$. This is given by
\begin{align}
    \Omega_{m}(\vec{\beta},\vec{\delta})&=\prod_{r=1}^{4}\prod_{i,j=-\infty}^{\infty}\prod_{k=0}^{m}(X^{\beta_{E^{i,j,k}_{d_{r}}}})^{\otimes {E^{i,j,m}_{d_{r}}}}(Z^{\delta_{O^{i,j,k}_{p_{r}}}})^{\otimes O^{i,j,m}_{p_{r}}}.
\end{align}
The initial state of the cluster, prior to measurement, is given by
\begin{align}
    &\rho_{CS} = \prod_{r=1}^{4}\prod_{i,j=-\infty}^{\infty}\prod_{k=0}^{m}(|+\rangle\langle+|)^{\otimes E^{i,j,k}_{p}}(|+\rangle\langle+|)^{\otimes O^{i,j,k}_{p_{r}}}
    \times(|0\rangle\langle0|)^{\otimes E^{i,j,k}_{d_{r}}}(|0\rangle\langle0|)^{\otimes O^{i,j,k}_{d_{r}}}.
\end{align}
Therefore letting
\begin{align}
    \mathcal{O}_{m} = [\sigma(\vec{\alpha},\vec{\gamma})\Omega_{m}(\vec{\beta},\vec{\delta})
    \prod_{i,j=-\infty}^{\infty}\prod_{l=1}^{4}\omega_{i,j,l}(\vec{\beta},\vec{\delta})]
\end{align}
\begin{align}
    &\tilde{\rho} = \mathcal{O}_{m} \cdot\rho_{CS}\cdot\mathcal{O}_{m}^{\dagger}
\end{align}
is the final state of the cluster after measurement, with measurements commuted to the final timeslice $t = m.$

\subsection{Extension to Fractalized Model}
We now consider the tensor network of the fractalized 3D RBH model described in Eq. \ref{eq19}. Since we are considering commuting measurement outcomes through qubit layers in the time direction, we need only worry about fractalization  in this dimension, which is specified by the function $h(w).$ 
Our $\sigma(\vec{\alpha},\vec{\gamma})$ operator is modified to the definition below, in order to account for additional qubits in the $w$ dimension
\begin{align}
    \sigma(\vec{\alpha},\vec{\gamma})&=\prod_{w=1}^{\infty}\prod_{l=1}^{4}\prod_{i,j=-\infty}^{\infty}\prod_{k=0}^{m}Z_{E^{i,j,k,w}_{p}}^{\alpha_{E^{i,j,k,w}_{p}}}X_{O^{i,j,k,w}_{p_{l}}}^{\gamma_{O^{i,j,k,w}_{p_{l}}}}.
\end{align}
As in the non fractalized case, we have an operator which comes from propagating  measurement outcomes through to $t = m$ from dual qubits on even time slices, and primal qubits on odd time slices. We need to extend this operator to these measurement outcomes for $w>0.$
For even time slices $t =2k$, primal qubits at $w=0$ which are entangled to dual qubits at $t = 2k + 1$, $h(w).$ For odd time slices $t = 2k+1$, dual qubits at $w  = 0$ that are entangled to primal qubits at $t = 2k+2$, $\bar{h}(w).$ 
We define an operator that accounts for the propagation of dual measurement outcomes through the lattice to $t = m$, which includes the effect of the additional connections between dual qubits due to fractalization.
This leads to the operator
\begin{align}
    \omega_{i,j,l}(\vec{\beta},\vec{\delta}) &= \prod_{w=0}^{\infty}\prod_{k=0}^{m-1}(X^{\beta_{E^{i,j,k,w}_{d_{l}}}})^{\otimes {E^{i,j,k',w}_{d_{l}}}:k'>k}
    (X^{\beta_{E^{i,j,k,w}_{d_{l}}}})^{\otimes O^{i,j,k',w}_{d_{r}}: O^{i,j,k,w}_{d_{r}}\in\mathcal{N}(E^{i,j,k,w}_{d_{l}}),k'>k}\\\nonumber
    &\prod_{b=1}^{n_{h}}(Z^{\delta_{O^{i,j,k,w}_{p_{r}}}})^{\otimes {O^{i,j,k',h^{(b)}(w)}_{p_{r}}}:k'>k}, 
\end{align}
where the final term in the product represent the effect of commuting the $Z^{\delta_{i,j,k,w}}$ measurement outcomes through the $CZ$ gates realising the fractalization.
Finally, we write down an operator representing the dual measurement outcomes having been commuted to $t=m$
\begin{align}
    \Omega_{m}(\vec{\beta},\vec{\delta})&=\prod_{w=0}^{\infty}\prod_{r=1}^{4}\prod_{i,j=-\infty}^{\infty}\prod_{k=0}^{m}(X^{\beta_{E^{i,j,k,w}_{d_{r}}}})^{\otimes {E^{i,j,m,w}_{d_{r}}}}(Z^{\delta_{O^{i,j,k,w}_{p_{r}}}})^{\otimes O^{i,j,m,w}_{p_{r}}}.
\end{align}

\twocolumngrid

\end{document}